\documentclass[a4paper,12pt]{article}
\usepackage{authblk}
\usepackage[utf8]{inputenc}
\usepackage[margin=0.75in]{geometry}
\usepackage{graphicx}
\usepackage{amsmath}
\usepackage{amsfonts}

\title{A review on the questions of spin and spin quantum correlations in the relativistic regime}
\author{Shrobona Bagchi}
\affil{Center for Quantum Information, Korea Institute of Science and Technology, Seoul, 02792, Korea.}
\date{}

\begin{document}
\maketitle

\begin{abstract} 

The majority of current understanding of the quantum correlations is in the field of non-relativistic quantum mechanics. To develop quantum information and computation tasks fully, one must inevitably take into account the relativistic effects. In this regard, the spin is one of the central tools. For this purpose, it is of paramount importance to understand and characterize fully the theory of spin in relativistic quantum information theory where the spin states act as qubit. This area is still far from being resolved. As a result, this article will explore the recent studies of the concepts of the spin and spin quantum correlations in inertial frames and some apparent paradoxes regarding this concept. We will mainly focus on the problem of characterizing the spin, reduced spin density matrices and spin quantum correlations in inertial reference frames and the apparent paradoxes involved therein. Another important aspect is the use of tools of quantum field theory to extend several concepts in non-relativistic domain to relativistic one. In this regard, we analyze the development of the theory of relativistic secret sharing and a correlation measure namely the entanglement of purification. \end{abstract}

\section{Introduction} 
Quantum correlations are an important part of modern quantum information theory  \cite{1,2,3}. This area is very well studied, developed and understood in the non-relativistic quantum information using non-relativistic quantum mechanics  \cite{4}. The study of these correlations have helped in the development and implementation of various quantum information processing protocols such as quantum teleportation, quantum cryptography and quantum secret sharing. However, this theory is not well understood and developed in relativistic quantum information theory that uses relativistic quantum mechanics or that uses quantum field theory \cite{5,6,7,8,9}. There are several less understood area that require careful analysis and resolution. We handpick a few such areas and explore in detail the fundamental and underlying issues, explore some fundamental and important quantum information task in relativistic quantum information and end with conclusions and potential future directions. These discussion inevitably fall in the section of quantum information in high energy physics, since the effect of relativistic velocity or acceleration become relevant in high energy physics. At the end, we also discuss how an important concept of entanglement of purification has been generalized in the area of holography, reflect on the conjecture and raise a few important issue regarding the operational significance of it in this area. These analysis shall also be helpful in the future space satellite based quantum communication.
\vskip 10 pt

As per our knowledge in quantum mechanics and relativistic quantum mechanics, the `spin' is usually thought of as an intrinsic `angular momentum' associated with the elementary particles like electrons, protons or other particles as such. It is understood to be purely a quantum mechanical property without any counterpart in classical physics. Spin in quantum mechanics is thought of as an existing `intrinsic' angular momentum of the particle and it is not due to the classical rotation of any internal component of the particle. Spin as a relativistic concept is still undergoing revisions to be understood to be in full glory and entirety. It is an essential part in majority of quantum information tasks in modern day quantum information and computation theory and applications. The important details start to emerge as one goes to inertial and non-inertial frames and several paradoxes and inconsistencies start to show in this arena in terms of its definitions, conceptualization and further generalizations such as the reduced spin density matrices in inertial frames. These observations point to the fact this is a partially developed concept and its development will lead to robust implementation of various quantum information processing tasks in relativistic regimes such as in future satellite based quantum communication. In the upcoming paragraphs, we visit some of these areas in short which explain later in details in the main sections later.
\vskip 10 pt
Spin qubit is a well understood concept in the non-relativistic quantum mechanics, It is understood very well through the implementation of the Stern-Gerlach experiment in the non-relativistic (low velocity which is much smaller than the speed of light in vacuum) limit. However, definition and understanding of the spin in the relativistic regime is underdeveloped till to this day. This has been analyzed in various works spread out in around the last decade and research on this is undergoing till today \cite{10,11,12,13,14,15, 16,17}. One of the origin of this difficulty lies in the phenomenon of the entangling of the spin degree of freedom with the momentum degree of freedom in the Lorentz-boosted reference frames. The second difficulty arises from the fact that for a quantum particle moving in a superposition of velocities as a quantum mechanical possibility, it is impossible for it to suddenly transition to its rest frame, wherein its spin is defined and understood properly. Possible remedy for these problems have been proposed by various authors. They eventually propose solution to this problem and ways of experimentally observing the relativistic features of the spin which then in turn promises to open up the possibilities of devising quantum information protocols using spin as a qubit in the special-relativistic regime.
\vskip 10 pt
Another very important and recurring problem in relativistic quantum information or  quantum information in high energy physics is the robust formulation of the reduced spin density matrix in the relativistic regime. Though the reduced density matrix for spin degree of freedom is well defined in non-relativistic quantum mechanics, its definition and formulation runs into problems when tries to simply extend this to relativistic regime. An apparent paradox involving the definition and formalism of spin density matrix in relativistic regime is given in \cite{11}. It was shown that a model for particle detection wherein a linear application of the Wigner rotations was applied to the state of a massive relativistic particle in a superposition of two counter-propagating momentum states, leads to a paradox. The paradoxical behavior is that the probability of finding the relativistic quantum particle at different positions depends on the reference frame, which is unwanted feature in the theory. A solution to the paradox was given there. According to the proposed solution,the authors argue that that we cannot in general linearly apply the Wigner rotations to a quantum state without considering the appropriate physical interpretation.
\vskip 10 pt
Again in similar vein, there is another problem similar in nature to the above. An open problem in the field of relativistic quantum information is whether entanglement and the quantitative  degree of violation of Bell’s inequalities for massive relativistic particles are dependent on the frames of reference or not. At the heart of this question lies the effect that spin degree of freedom gets entangled with the momentum degree of freedom at relativistic regime. In a more advanced work, the authors here show that the Bell’s inequalities for a pair of quantum particles can be maximally violated in a `special-relativistic regime', even without any post-selection of the momentum of the particles, shown via the use of the novel methodology of quantum reference frames. The authors in claim that the use of the quantum reference frames allow them to transform the problem to the rest frame of a particle, whose state can be in a superposition of relativistic momenta from the viewpoint of the laboratory frame of reference. In this work, the authors work with several problems of defining the spin density matrix in relativistic regime and show that when the relative motion of two particles is non-collinear, the appropriate measurements for violation of Bell’s inequalities in the laboratory frame involve  the “coherent Wigner rotations”. In this work, the authors also show that the degree of violation of Bell’s inequalities is independent of the choice of the newly introduced and defined quantum reference frames, which is a desired feature in a theory. 
\vskip 10 pt
After the full description of the existing line of work of the subject of spin density matrices in relativistic reference frames, we turn our attention to the use of tools of quantum field theory in implementing quantum information protocols that are in the high energy regime \cite{18,19,20}. In this respect we review a protocol of relativistic quantum secret sharing in relativistic regime. Here, the authors develop a quantum secret sharing protocol that relaxes usual assumptions and consider the effects due to the accelerating motion. 
\vskip 10 pt
Another developing area of research is the various extensions and further development of various quantum information correlation measures via quantum field theory and conformal field theory  \cite{21,22}. One of such a definition is the entanglement of purification which defines the total correlation measure for a quantum particle in an operational way \cite{23}. Entanglement wedge cross section has been developed to account for the counterpart of entanglement of purification in quantum field theoretic and conformal field theoretic terms. It has been termed as the Holographic Entanglement of Purification by Tadashi Takayanagi and Koji Umemoto  \cite{23,24,25}. They suggest that it is a holographic counterpart of entanglement of purification, which measures a bipartite correlation in a given mixed state quantum state defined on an operational basis. We point out in our coming sections that a similar operational footing in the quantum field theoretic terms will be a promising area of research in future.

\section{On the questions of spin and spin quantum correlations in relativistic quantum mechanics and relativistic quantum information}

In this section, we review in short a few key concepts needed to understand the basic analysis of spin in quantum mechanics and the problems associated with it for its formulation in the relativistic regime. These concepts include those of reference frames as defined in theory of special relativity. Any physical system is defined using a set of coordinates that completely specifies its reference frame. From the principles of special relativity, we know that all physical laws hold the same way in all reference frames. The laws of physics transform covariantly in between the different reference frames. However, there is typically a reference frame which is the most convenient to use, where the system is at rest. This reference frame is called the rest frame of the physical system. These concepts of the classically defined reference frames as in classical physics are very well understood in the absence of intricacies of certain quantum mechanical phenomenon. However, when we start to analyze some quantum properties in detail by using the traditionally defined reference frames, we are faced with some problems. In this respect it was shown in that when the external degrees of freedom for example the momentum of the physical system are in a quantum superposition with respect to the laboratory frame of reference, no classical reference frame transformation, as prescribed by the special theory of relativity, can map the description of physics from the laboratory to the rest frame \cite{12}. This area of physics is not understood well enough. Subsequently the concept of quantum reference frame was  introduced and leveraged to give an operational footing to the concept of spin in relativistic quantum mechanics \cite{12}. It was claimed in \cite{12} that such a formulation is able to solve some of the paradoxical features related to transformation of spin in relativistic quantum mechanics. There are several claims along this direction by various authors. As a result proper experimental verifications are needed to settle the correct theory for spin in relativistic quantum mechanics and relativistic quantum information theory. These concepts are explained later in the later paragraphs.
\vskip 10 pt
One of the single-most important concept include that of the spin in non-relativistic quantum mechanics defined operationally via the Stern-Gerlach experiment. The spin of a Dirac spin-1/2 particle is defined by the $2 \times 2$ Pauli matrices $\sigma_i (i = 1, 2, 3)$. The Pauli matrices, together with the unit matrix generate an irreducible representation of the $SU(2)$  group. It is well known that the spin operator of a non-relativistic spin-1/2 particle in quantum mechanics is very well understood. Also, we know that there is a clear correspondence between quantum-mechanical operators and classical variables in non-relativistic quantum mechanics. This correspondence exists for all operators such as the position, momentum, and angular momentum etc. However, in contrast to this observation, the connection between the quantum-mechanical operators and classical variables in relativistic Quantum Mechanics is much more subtle and complex. 
\vskip 10 pt
In this respect, we review the concept of spin in relativistic quantum mechanics from different perspectives, provided by various authors until now.  They involve the analysis of an apparent paradox caused by linear Wigner rotation and the quantum reference frames. These are presented below.

\subsection{Apparent paradox of Wigner rotations for relativistic effects of spin of quantum particle}

The intricacies of the conceptual and analytical foundation of spin of relativistic quantum particle was presented in \cite{11}. In their analysis, the authors presented an apparent paradox involving the definition and formalism of spin density matrix in super-relativistic regime. It was shown there that a method of particle detection in combination with a linear Wigner rotations, which then corresponds to momentum-dependent changes of the particle spin owing to the fact that spin and momentum degrees of freedom get entangled in relativistic scenario under Lorentz transformations, applied to the state of a massive relativistic quantum particle in a superposition of two different momentum states leads to a paradox. The paradoxical feature is that the probability of finding the relativistic quantum particle at different positions depend on the reference frame, which should not be the case. As a solution to the paradox which is also simple, the authors suggested that one cannot in general linearly apply the linear Wigner rotations to a quantum state without considering the appropriate physical interpretation about it.
\vskip 10 pt 
We sketch out the main steps of their analysis here. The initial state taken is this case is of the following form 
\begin{align}
|\Psi\rangle = \frac{1}{\sqrt{2}}[|p\hat{y},Z\rangle-|p\hat{y},-Z\rangle],
\end{align}
in the frame $S_0$. Here, $ |p, \pm Z \rangle $ represents a state for the particle with 4-momentum $(p^0,\vec{p})$, and spin state pointing in the $\pm z$ direction, being the eigenvector of the Pauli matrix $\sigma_z$ with eigenvalue $\pm 1$ with reference to $S_0$. Here, the authors have used Wigner’s definition for spin and they have taken $c=1$. Therefore, from the perspective of the rest frame, the quantum particle is in a superposition of momentum of opposite directions. Using proper algebra it was found out by the authors that the probability density of finding the relativistic quantum particle around position $y$ obeys the following expression
\begin{align}
p_0(y) \propto \sin^2(\frac{py}{\hbar})
\end{align}
If one then makes a change of reference frame to a frame that moves with velocity $\beta z$ in relation to $S(0)$, each momentum component of the state $|\Psi\rangle$ undergoes a different spin transformation. The spin and the momentum degrees of freedom get entangled when linear Wigner rotations are applied in a non-trivial way. See \cite{11} for details.  In the reference frame $S_1$, the momentum state of the particle changes from before, though the $y$ component remains the same. In this work, the authors then analyze the expressions about the $y$ dependence of the particle wavefunction and do not analyze for $z,x$ direction for simplification without any loss of generality. With some reasonable approximation, the probability expressions found with respect to reference frame $S_1$ is given by the following 
\begin{align}
p_1(y)\propto \cos^2(\frac{\phi}{2})\sin^2(\frac{py}{\hbar})+\sin^2(\frac{\phi}{2})\cos^2(\frac{py}{\hbar}),
\end{align}
where the angle $\phi$ is related to the boost parameter for reference frame $S_1$ \cite{11}. This new expression for the probability therefore points to a paradox that has crept in the calculation in between, done in the traditional way. The authors in \cite{11} argues that this is a paradox since the probability of finding the particle around some position should depend on the reference frame. As a result, this paradox therefore points towards a deficiency of the current state of art of the theory of spin in quantum mechanics. Thereafter the authors in \cite{11} test this observation in a different way via the use of quantum measurements using detectors. They consider measurements of the particle position using a detector that, by construction, responds only to the charge or the mass of the particle, but does not in any case depends on its spin. Using this formalism and again placing a small number of reasonable approximations that are relevant in experimental setup, they again show the discrepancy in the expressions for probability expressions as calculated with respect to different reference frames.
\vskip 10 pt
To summarize, they have shown that the application of the momentum-dependent linear Wigner rotations to the quantum state of a massive relativistic particle in a superposition of counter-propagating momentum states along with a model for particle detection leads to a paradox, since the probability of finding the particle at different positions would depend on the reference frame. Considering the physical implementation of the quantum state, they have discussed that the Wigner rotation depends on the preparation method, such that, with a change of the reference frame, the spin transformation of a quantum state in a superposition of different momenta is not exactly equivalent to the linear application of the momentum-dependent Wigner rotation to each momentum component of the state. This they say solves the apparent paradox. Their work, together with a few previous works on the subject, show that relativistic quantum transformations cannot in general be computed only by following a mathematical procedure as in the traditional literature of relativistic quantum mechanics. The authors argue that the physical meaning of the transformations must always take precedence before their application in that physical scenario.
\vskip 10 pt
Though they have proposed the above solution to the apparent paradox, however they also stress on the fact that their solution may not be the only viable solution to this apparent paradox. It  may be possible that by modeling the particle detection by some more complicated scheme the paradox could be solved keeping the linearity property of the Wigner rotations. Consequently different solutions were proposed by different authors to this problem, one of the main contenders in the name of quantum reference frames \cite{12} is discussed in the next sections. However, it is important to note that the way to settle and reach a proper solution to this apparent paradox needs experimental verifications under different conditions and approximations.

\subsection{Relativistic Stern Gerlach Experiment and Quantum Reference Frames}

To provide a consistent description of the relativistic effects on the spin of a quantum particle, more theories were proposed, one of which is based on the relatively recent proposals of quantum reference frames. Along this line, the relativistic treatment of the Stern-Gerlach experiment was proposed and was termed as the relativistic Stern-Gerlach experiment \cite{12}. The theory of the quantum reference frames was evoked there to give an operational interpretation to the spin in relativistic quantum mechanics. It was noted in \cite{12}, that when the particle has relativistic velocities, the spin degree of freedom transforms in a momentum-dependent way, as was noted by previous authors. Then, if a standard Stern-Gerlach measurement is performed on a particle in a pure quantum state moving in a superposition of relativistic velocities, the operational identification of the spin fails, because it was shown in \cite{12} that no orientation of the Stern-Gerlach apparatus returns an outcome with unit probability. The question that then arises is whether it is possible to find ‘covariant measurements’ of the spin and possibly momentum, which predict invariant probabilities in different Lorentzian reference frames for the case of a quantum relativistic particle moving in a superposition of velocities \cite{12}. This is therefore an alternate solution to the solution proposed in \cite{11} as described in the previous section. If such measurements are possible to construct, then it would be possible to map the description of spin in the rest frame of the particle to the frame of the laboratory in an unambiguous way. This would enable one to derive the corresponding observables to be measured in the laboratory frame to verify the correct theory of the spin of a relativistic quantum particle.
\vskip 10 pt
The trial for finding such covariant measurements that preserve probability values in different reference frames is motivated by the potential applications where the spin degree of freedom is used as a qubit, to encode and transmit quantum information in the relativistic regime. Earlier protocols as such therefore are no longer valid in a relativistic context. This severely constraints and undervalues the wide range of applicability of techniques involving spin as a quantum information carrier in the relativistic regime. It is then important to explore possible methods which can overcome this limitation. In the context of relativistic quantum information, this question has been extensively discussed in relation to Wigner rotations and has been related to the problem of identifying a covariant spin operator \cite{11,12,13,14,15}. A variety of relativistic spin operators have been proposed till date. Some of them are called the Frenkel, the Pauli-Lubanski, the Pryce, the Foldy-Wouthuysen, the Czachor the Fleming the Chakrabarti, and the Fradkin-Good spin operators. 
\vskip 10 pt
To remedy the above problems, the authors in \cite{13} use the concept of ‘superposition of Lorentz boosts’ which allow them to make the relativistic quantum particle “jump” into the rest frame even if the particle is not in a momentum eigenstate but in a quantum state with a superposition of momentum in general. It is well known that in the rest frame, the spin observables satisfy the $SU(2)$ algebra and are operationally defined through the famous Stern-Gerlach experiment. The authors in \cite{13} aim to make this same concept work in inertial reference frames. In the work \cite{13}, the authors transform the set of spin observables in the rest frame to an isomorphic set of observables in the laboratory frame. The transformed observables are in general entangled in the spin and momentum degrees of freedom as expected. The new set fulfills the $SU(2)$ algebra again and is operationally defined through an experiment which the authors label as the ‘relativistic Stern-Gerlach experiment’. In this experiment, the authors construct the interaction term and the measurement term between the spin-momentum degrees of freedom and the electromagnetic field in the laboratory frame which gives the same probabilities as the Stern-Gerlach experiment in the rest frame, as desired and stated earlier in the paragraph. This set of observables in the laboratory frame allows the authors to partition the total Hilbert space into two subspaces corresponding to the two outcomes which can be termed as “spin up” and “spin down”. Hence, with techniques of the quantum reference frames, the relativistic spin states can effectively be used to  construct a qubit state in an operationally well-defined way, as claimed by the authors in \cite{13}. Thus the quantum reference frames and the relativistic Stern-Gerlach experiment promise to be a robust candidate representing the theory of intrinsic spin of relativistic quantum particles and its transformations in between different reference frames. However, the correctness of this theory is still open to experimental demonstrations to be established fruitfully.
\vskip 10 pt
With the above background in mind, we now move on to the description of the relativistic Stern Gerlach experiment as discussed in \cite{12}. 
One considers an experiment performed in the laboratory reference frame referred to as C. One allows the particle to have any quantum state, and in particular to move in a superposition of momenta. This condition implies that there is a non-classical relationship between the two reference frames. This means that the rest frame A and the laboratory frame C are not related by a standard boost transformation as in classical special relativity. The authors in this work \cite{12} implement a method to generalize the boost transformation to this case of the relativistic quantum particle.  As for the coordinates for the mathematical analysis, $x$ and $z$ are used to describe the external degrees of freedom of particle $A$ and the intrinsic spin degrees of freedom as $\tilde{A}$ of the relativistic quantum particle. The state of the particle time $t = 0$ is taken to be the following
\begin{align}
|\Psi\rangle=\cos\theta |\psi^+\rangle+\sin\theta|\psi^-\rangle, 
\end{align}
where again we have the following definitions
\begin{align}
|\psi^\pm\rangle=|\psi_z\rangle_A|\phi_x^\pm\rangle_{A\tilde{A}}, 
\end{align}
is a division of the total wave function into components in $x$ and $z$ directions in rest frame and lab frames as per the notation. It is assumed that the motion along $z$ direction is non-relativistic, without any loss of generality. Writing these wave functions in terms of superposition of momentum states, we have the following expressions
\begin{align}
|\psi_z\rangle_A=\int dp_z \psi_z |p_z\rangle_A, 
\end{align}
with $\psi_z$ denoting Gaussian wave functions in the momentum variable $p_z$ centered around $p_z=0$ and standard deviation $s_z$. The other component is denoted as the following 
\begin{align}
|\phi_x^\pm\rangle_{A\tilde{A}}=\int d\mu(p_x) \phi_x |p_x,\Sigma_{p_x}^\pm\rangle_{A\tilde{A}}, 
\end{align}
where $\phi_x^\pm$ is a general wavepacket expression and $|\Sigma_{p_x}^\pm\rangle_{A\tilde{A}}$ are the eigenvectors of the operator obtained via Lorentz boost and Pauli-Lubanski operator as defined in \cite{12},  with eigenvalues $\pm 1$.  In the laboratory frame, it is possible to define the observables corresponding to the spin operators in the rest frame by transforming the spin, as defined in the rest frame, with a quantum reference frame which then correspond to the transformation. They are called the manifestly covariant Pauli-Lubanski spin operator. Now after this, the authors engineer a Hamiltonian with the following interaction term Hamiltonian in the laboratory frame as follows
\begin{align}
H_{int}=\mu B_z \xi,
\end{align}
where, $B_z=B^0_z-\alpha z$ . $\xi$ is the term containing the components which are obtained using the components of the manifestly covariant Pauli Lubanski spin operator modified with parameters dependent on the boost parameters, as defined in \cite{12}. Let us now see how the operators $\xi$ came into the picture. The authors in \cite{12} note that in the laboratory frame C, when the particle A is in a superposition of momentum states, no spin measurement in a standard Stern-Gerlach experiment gives a result with probability one, because of following two reasons: the spin and momentum are entangled, and the relation between the laboratory and the rest frame is not a classical special relativistic reference frame transformation. In order to devise such measurement that will give consistent probability values in all reference frames as in the rest frame, the authors in \cite{12} note that in the laboratory frame, it is possible to define the observables corresponding to the spin operators in the rest frame by transforming the spin, as defined in the rest frame, with a Quantum Reference Frame transformation, the expression of which is then derived as $\xi$, the details of which can be found in \cite{12}.
\vskip 10 pt
Now let us look at the dynamics of the relativistic quantum particle due to the Hamiltonian as described above. The Hamiltonian term is an appropriate interaction Hamiltonian containing a magnetic field in the $z$ direction in the laboratory frame. The state is then evolved with the action of this Hamiltonian and its form is written down appropriately in the interaction picture as a function of time. It was the shown in that under the effect of the interaction with the magnetic field, the gaussian wavepacket along $z$ gets split into two wavepackets, moving in opposite directions according to the state of the spin. After this appropriate projection operators are applied to the wave packet and probabilities of obtaining spin “up” or spin “down” as value is obtained denoted by $p^\pm$. It was shown via this calculation in \cite{12} that for a time when the two wavepackets become distinguishable, the probabilities for obtaining up and down spins are found out to be $\cos^2\theta$ and $\sin^2\theta$, when irrevalent terms are neglected under appropriate limit and subsequent approximation. The authors claim that in this way, one can solve the problem of ambiguity of finding the correct expressions for probabilities in rest frames in the relativistic Stern-Gerlach experiment.
\vskip 10 pt
In this work, the authors claimed to have provided a correct operational description of the spin of a special-relativistic quantum particle which has been elusive for a while. Such operational description was initially difficult to obtain with standard traditional treatments due to the combined effect of special relativity and quantum mechanical properties, which makes the spin and momentum entangled and an impossibility of jumping to the rest frame with traditional tools. To remedy this problem, the authors in have introduced the concept and mathematical characterization of ‘superposition of Lorentz boosts’ transformation to the rest frame of a quantum particle, moving in a superposition of relativistic velocities from the point of view of the laboratory reference frame.  As a result of their analysis based on the quantum reference frames, probabilities obtained in the relativistic Stern-Gerlach experiment are shown to remain the same in the rest frame and in the laboratory frame, which was challenging task to accomplish before. This approach is relatively new with respect to some earlier approaches and proposed  theoretical remedies. However, it should be emphasized that the theoretical treatment offered in this work is yet to undergo several experimental checks in different limits, experimental conditions and relevant approximations to be accepted and established as a correct theory for relativistic effects in spin of a quantum particle. 

\subsection{ Other effects related to relativistic treatment of spin of a quantum particle}

There are several other effects associated to the correct description of spin in relativistic quantum mechanics. An open question in the field of relativistic quantum information is the question of invariance of a measure of entanglement and/ or the quantitative degree of violation of Bell’s inequalities for massive relativistic particles in different frames of reference. Such questions can also be extended similarly to other quantum information theoretic correlation measures. Likewise as before, again at the core of this dilemma is the effect that spin gets entangled with the momentum degree of freedom at relativistic velocities. In \cite{13}, the authors claim to show that the Bell’s inequalities for a pair of particles can be maximally violated in a special-relativistic regime, even without any post-selection of the momentum of the particles, again via the use of the concept of quantum reference frames. They specifically show that, when the relative motion of two particles is non-collinear, the optimal measurements for violation of Bell’s inequalities in the laboratory frame involve “coherent Wigner rotations” \cite{13}. Thus, they also touch upon a debated concept of appropriate application of Wigner rotations in in this physical set up. In this formalism, the authors also show that the degree of violation of Bell’s inequalities is independent of the choice of the quantum reference frame, which is a desired effect. As a result, this work attempts to settle some important open questions involving the fundamental concepts of spin and relativity. However, for it to be established to be a correct theory or otherwise, several experimental checks are needed to be performed in a consistent and reproducible way. As a result, experimental proposals to test out these theories in the laboratory under different conditions pose to be the next important steps in full development of this area of research and enquiry.

\subsection{ Experimental efforts to test relativistic theory of spin of quantum particle}

Experimentally there have been many efforts recently to measure the quantum spin correlations of elementary particles. One of those proposed experiments involve the study of quantum spin correlations of relativistic electron pairs for testing the non-locality of relativistic quantum mechanics. Finding the right expression and formalism for spin and spin quantum correlations in relativistic quantum mechanics is an important direction of research both from the perspective of space communications as well as testing the fundamentals of quantum physics and quantum gravity. Here we report on two attempts at experiments to measure the spin quantum correlations in relativistic scenarios. 
\vskip 10 pt
An experiment investigating the quantum spin correlations of relativistic electrons is reported here \cite{16}. The project presented in \cite{16} tries to make the first measurement of the quantum spin correlation function for a pair of massive relativistic particles. This measurement is claimed to be the first attempt to verify the predictions of relativistic quantum mechanics in the domain of spin correlations. This is an interesting direction of research since it has the capability of settling competing theories of spin quantum correlations experimentally or even point out unknown deficiencies in the foundations of relativistic quantum mechanics. As per their description of the proposed experiment, the measurement is carried out on a pair of electrons in the final state of Moller scattering. The measurement attempts to measure correlations of spin projections on chosen directions for the final state pair after the complete evolution via the chosen dynamics. The detector consists of two Mott polarimeters, in which the spins of both Moller electrons are measured simultaneously. However, the results have not yet been linked to the theoretical predictions of the quantum reference frames or other competing theories of spin in relativistic quantum mechanics. This remains a promising future direction of research.
\vskip 10 pt
Another direction of research has been the study of quantum spin correlations of relativistic electron pairs for the purpose of testing non-locality of relativistic quantum mechanics. The theory developed along this direction has been discussed in the previous sections. Therefore an experiment directed at testing the predictions offered by the theory for example as offered by the quantum reference frames will be extremely helpful in settling the correct theory of relativistic effects of quantum spin among many competing theories. This will help advance the understanding of fundamental theory in nature. This project is supposed to be a Polish-German project QUEST that will aim at studying relativistic quantum spin correlations of the Einstein-Podolsky-Rosen-Bohm type, through appropriate measurements and the corresponding probabilities for relativistic electron pairs. This experiment will also use the Moller scattering method and Mott polarimetry technique. 

\section{Quantum Information With Quantum Field Theory: Relativistic Quantum Secret Sharing }

In the previous section, we discussed the fundamental question of the concept of spin in relativistic quantum mechanics. We discussed how the concept of spin is an interesting topic, and how its understanding and unraveling of exact nature and experimental verification will help in development of intricate quantum technologies that use the spin as qubit in quantum information processing tasks. However, though it is important to understand the concept of spin in relativistic quantum mechanics, one can also use the concept of quantum information processing tasks in relativistic quantum information using tools from quantum field theory. One of these approaches has been to use cavity dynamics for the case of non-inertial motion in general. In this section, we report on the protocol of quantum secret sharing in the relativistic setting. Here we focus more on the theory of a specific quantum information task called relativistic quantum secret sharing using tools from quantum field theory.
\vskip 10 pt
In (2,3)-threshold quantum secret sharing, the 'dealer' which is one pf the parties taking part in the protocol, encodes the quantum secret in three quantum shares in a localized manner. The authors in \cite{12} use the framework of accelerating cavities for this purpose, as it is a suitable choice to study the effect of non-uniform motion on localized quantum fields. Accelerating cavities are popular to study the relativistic effects on quantum information protocols. However, the authors in \cite{12} develop a different approach from the approach of accelerating cavities. They formulate the evolution of the quantum field inside an accelerating cavity as a bosonic quantum Gaussian channel which they then use to include the effects of non-uniform motion of the quantum shares. 
\vskip 10 pt
The authors in \cite{20} focus on a relativistic variant of a (2,3)-threshold quantum secret sharing protocol. In the relativistic protocol presented in \cite{20}, similar to the non-relativistic case, a dealer encodes the quantum secret into several quantum shares and distributes them to all the players. In this set up, the players are all located at different regions in the Minkowski spacetime and the dealer and the players are all stationary. Under such circumstances, during the dealer’s distribution, the quantum shares experience non-uniform motion (non-inertial), as they are transmitted to spacetime points in the future light cone of the dealer. Then, a subset of players within the access structure collaborate to retrieve the quantum secret by sharing their individual shares. However, to reach the same spacetime point, the shares go through phases of accelerating and decelerating motion while being transmitted, rendering the dynamics to be non-inertial in general, in contrast to the special-relativistic regime described in the previous sections. The authors investigate how the non-inertial motion of the shares affects the fidelity of the quantum secret sharing protocol. The authors in this work claim to have solved continuous-variable quantum secret sharing wherein the quantum shares move non-uniformly in Minkowski spacetime. The tools used in this approach mainly comprise of the tools developed in continuous variable quantum information such as the formalism of the Gaussian quantum channels, dynamics of quantum field inside cavity \cite{20}. The authors do not use spin as the qubit in this relativistic scenario and yet are able to implement the relativistic quantum secret sharing protocol mainly via using the formalism of quantum field inside cavity which may themselves be in non-inertial motion \cite{20}. The authors in \cite{12} specify that they use the framework of Gaussian quantum information to write the evolution and dynamics of the quantum field inside the cavity central to their implementation of the protocol of quantum secret sharing, as a Gaussian quantum channel, in the non-inertial regime. They use this channel to study the effect of non-inertial motion of the shares on the fidelity of the quantum secret sharing. As a result, this work shows that various methods can be utilized to study the effects of non-inertial motion, without directly referring to the transformation of spin in super-relativistic regime.

\section{Entanglement of purification and entanglement wedge cross-section }

In this section, we discuss the relativistic quantum information from another angle. We discuss the quantification and characterization of correlation measures used in non-relativistic quantum information theory using quantum field theory in relativistic regime. The quantity that is usually used is the entanglement entropy. The entanglement entropy is used to quantify the entanglement of the pure quantum states.  Using the AdS/CFT correspondence, the entanglement entropy has a holographic counterpart given by the area of minimal surface \cite{25}. This characterization therefore provides a relationship between spacetime geometry and quantum entanglement
\vskip 10 pt
The entanglement is a pure quantum correlation. However, the mixed quantum states contain both the classical and quantum correlation together contained in the total correlation of the quantum state. Now, a usual measure of the total correlation of a mixed quantum state is the mutual information. However, the justification of the mutual information as a measure of total correlation is sometimes questioned. Also, there were attempts at separating the quantum correlation from the classical correlation in the quantum mutual information via the introduction of the measure of quantum correlation called quantum discord. Another approach was proposed to quantify the total correlation in a quantum state via the transformation of the entanglement of Bell pairs via local operation and classical communication. This measure is called the entanglement of purification. Its many properties were studied in \cite{24}. Later a holographic quantity was proposed as a counterpart of this entanglement of purification using the concept of entanglement wedge cross-section, as a conjecture. This quantity is called the holographic entanglement of purification. The definition of entanglement of purification in non-relativistic quantum mechanics is given as follows. If we have a given mixed quantum state $\rho_{AB}$, then at first we purify the mixed quantum state $\rho_{AB}$ to $|\Psi\rangle_{AA'BB'}$ and then, the entanglement of purification of the mixed quantum state $\rho_{AB}$ is given as the follows 
\begin{align}
E_P(\rho_{AB})=\min_{A'B'} E_f(AA':BB')
\end{align}
In the work of holographic entanglement of purification, the authors have considered the quantity $E_W$ defined as the minimal cross section of entanglement wedge in AdS/CFT. The authors have shown there that they have observed that its properties actually coincide with those of the quantity called entanglement of purification, which measures total correlation between two subsystems for a mixed quantum state that comprises of both classical and quantum correlations. Based on their observations and calculations on the entanglement wedge cross-section, the authors have conjectured that Entanglement wedge coincides with entanglement of purification in holographic conformal field theories. The authors also have given a heuristic argument for this identification. Open question remains whether this conjecture is true or not. Several works are ongoing to check this conjecture in mathematical terms.
\vskip 10 pt
Another very important and promising open direction along this line of research is finding an operational interpretation of the definition of entanglement wedge cross section in a similar way the entanglement of purification was motivated and operationally found in non-relativistic quantum mechanics. The entanglement of purification was motivated operationally in terms of the conversion of the purely quantum correlation called entanglement in the maximally entangled state called the Bell pairs into the total correlation measure in the arbitrary quantum state using local operation and classical communication (LOCC). Therefore, to settle the conjecture another direction could be to try to find an operational footing of the entanglement wedge cross-section as well. 

\section{Conclusions}

In this section, we summarize what we have discussed in the previous sections and open questions. Developing the fundamentals of understanding of nature and the natural phenomenon with a robust mathematical construct and therein subsequent reproducible experimental verifications has been one of the strongest pillars of physics as we know it today. Many technological applications have stemmed out from the robust structure of physical phenomenon developed in theories in physics. Thus, we can imagine that such further developments will open up immense possibilities in future technologies that have high potential for finding solutions for persistent problems in lives of people and such. Thus it is clear that from a point of view of understanding of nature, technological developments and even resource allocation, development of fundamental theories of nature is an area of research that hold immense potential.

In view of the above motivation, we have covered here a few aspects of physics that are important for the development quantum information theory in relativistic regime, i.e., in high energy physics where it is paramount to use the relativistic effects via relativistic quantum mechanics and quantum field theory. It is well known that the concept of spin is an ill understood concept in relativistic quantum physics. Prior approaches to spin in quantum physics has been discussed and some recent promising approaches have been presented. It has also been discussed how a robust formulation of the concept of spin is crucial for the development of quantum information in high energy physics. We also presented a review of the problem of spin quantum correlations and have presented the resources in scientific literature that are trying to verify this experimentally in recent years. This issue is still unresolved and a consistent resolution of this has the potential to open doors for the applications of relativistic quantum information and extension of them in space based quantum technologies.

With the above open problem in mind, we also take note of the fact that tools of quantum field theory can be leveraged to again develop quantum information protocols in relativistic quantum information in high energy arena, especially in the regime of non-inertial motion. Such a protocol of relativistic quantum secret sharing has been discussed in detail. Similar techniques can be leveraged to develop such protocols further in high energy physics.

In the last section of this chapter we have covered a section of the development of definition of a total correlation measure in the language of quantum field theory and conformal field theory. The concept of entanglement of purification which is an active and open area of research has been discussed here. Another open area of research has been pointed out in this arena which has high potential of development in future. This area is the operational characterization of the quantum correlation measures defined in terms of tool as in quantum field theory and conformal theory. This area of enquiry is based on the question of experimental verification in tabletop setups. This can be an active area of research in future that might be challenging yet highly promising and with potential for rich dividents.

\section*{Acknowledgments}

SB acknowledges funding from Korea Institute of Science and Technology. S. B. acknowledges support from the National Research Foundation of Korea (2020M3E4A1079939, 2022M3K4A1094774) and the KIST institutional program (2E31531).


\begin{thebibliography}{99}

\bibitem{1} Gerardo Adesso, Thomas R. Bromley, Marco Cianciaruso, Measures and applications of quantum correlations. J. Phys. A: Math. Theor. 49, 473001 (2016).

\bibitem{2} Quantum entanglement, R Horodecki, P Horodecki, M Horodecki, K. Horodecki: Rev. Mod. Phys. 81, 865,  (2009).

\bibitem{3}K Modi, A Brodutch, H Cable, T Paterek, V Vedral, The classical-quantum boundary for correlations: Discord and related measures: Rev. Mod. Phys. 84, 1655, (2012).

\bibitem{4}Ryszard Horodecki, Quantum information: Acta Phys. Pol. A 139, 197 (2021).

\bibitem{5}R B Mann and T C Ralph, Relativistic quantum information, Classical and Quantum Gravity, \textbf{29},  22 (2012).


\bibitem{6}I. Fuentes, Diversities in Quantum Computation and Quantum Information, Lecture series in quantum information, 107-147, (2012), World Scientific.

\bibitem{7}Fabrizio Tamburini (Editor), Ignazio Licata (Editor), Relativistic Quantum Information, Entropy, MDPI books (2020).

\bibitem{8}E Tjoa, K Gallock-Yoshimura, Channel capacity of relativistic quantum communication with rapid interaction: Phys. Rev. D 105, 085011 (2022).

\bibitem{9}E Tjoa, Quantum teleportation with relativistic communication from first principles, Phys. Rev. A 106, 032432 (2022).

\bibitem{10}PL Saldanha, V Vedral, Spin quantum correlations of relativistic particles, Phys. Rev. A 85, 062101 (2012).

\bibitem{11}LF Streiter, F Giacomini, Č Brukner, Relativistic Bell test within quantum reference frames, Phys. Rev. Lett. 126, 230403 (2021).

\bibitem{12}M Mikusch, LC Barbado, Č Brukner, Transformation of spin in quantum reference frames,  Phys. Rev. Research 3, 043138 (2021).

\bibitem{13}F Giacomini, E Castro-Ruiz, Č Brukner, Relativistic quantum reference frames: the operational meaning of spin, Phys. Rev. Lett. 123, 090404 (2019).

\bibitem{14}Daniel R. Terno, Two roles of relativistic spin operators, Phys. Rev. A 67, 014102 (2003).

\bibitem{15}Liping Zou, Pengming Zhang, and Alexander J. Silenko, Position and spin in relativistic quantum mechanics, Phys. Rev. A 101, 032117 ( 2020).

\bibitem{16} Marta Wlodarczyk, Paweł Caban, Jacek Ciborowski, Michal Dragowski, and Jakub Rembielinski: Quantum spin correlations in Moller scattering of relativistic electron beams, Phys. Rev. A 95, 022103 (2017).

\bibitem{17} Jacek Ciborowski, Pawel Caban, Michal Dragowski, Joachim Enders, Yuliya Fritzsche, Artem Poliszczuk, Jakub Rembielinski and Marta Wlodarczyk, A project to measure quantum spin correlations of relativistic electron pairs in Moller scattering: EPJ Web of Conferences 164, 01004 (2017), EDP Sciences.

\bibitem{18}T. Rick Perche and Eduardo Martín-Martínez, Role of quantum degrees of freedom of relativistic fields in quantum information protocols : Phys. Rev. A 107, 042612, (2023).

\bibitem{19}Dionigi M T Benincasa, Leron Borsten, Michel Buck and Fay Dowker: Quantum information processing and relativistic quantum fields:  Class. Quantum Grav. 31 075007(2014).

\bibitem{20} David Edward Bruschi, Antony R Lee and Ivette Fuentes, Time evolution techniques for detectors in relativistic quantum information: J. Phys. A: Math. Theor. 46 165303 (2013).

\bibitem{21} Pasquale Calabrese and John Cardy, Entanglement entropy and quantum field theory, Journal of Statistical Mechanics: Theory and Experiment, P06002, Volume 2004 (2004).

\bibitem{22} Tadashi Takayanagi, Entanglement entropy from a holographic viewpoint: Class. Quant. Grav. 29, 153001 (2012).

\bibitem{23}Mehdi Ahmadi, Ya-Dong Wu, and Barry C. Sanders, Relativistic (2,3)-threshold quantum secret sharing:  Phys. Rev. D 96, 065018 (2017).

\bibitem{24} Shrobona Bagchi and Arun Kumar Pati, Monogamy, polygamy, and other properties of entanglement of purification: Phys. Rev. A 91, 042323( 2015).

\bibitem{25}Koji Umemoto and Tadashi Takayanagi, Entanglement of purification through holographic duality: Nature Physics volume 14,  573–577 (2018).

\end{thebibliography}
\end{document}